
\def\SU#1{{\rm SU}(2)_{#1}}
\def\sl#1{\hat{sl}(2)_{#1}}
\def\slq#1{U_{#1}(sl(2))}
\def\pno{\par\noindent}
\def\ref#1{$^{#1}$}
\def\aa{\alpha}
\def\bb{\beta}
\def\ga{\gamma}
\def\De{\Delta}
\def\la{\lambda}
\def\2pi{2\pi i}
\def\M#1{{\rm M}_{#1}}
\def\N#1{{\rm N}_{#1}}
\normalbaselineskip = 12 pt
\magnification = 1200
\hsize = 15 truecm \vsize = 22 truecm \hoffset = 1.0 truecm
\rightline{LANDAU-93-TMP-3}
\rightline{April 1993}
\rightline{Submitted to Int. J. Mod. Phys. A}
\rightline{hep-th/9304116}
\vskip 2 truecm
\centerline{{\bf COSET CONSTRUCTION OF MINIMAL MODELS}}
\vskip 2 truecm
\centerline{M. YU. LASHKEVICH\footnote{$^*$}{ This work was supported,
in part, by Landau Scholarship
Grant awarded by Forschungszentrum J\"ulich, and by Soros Foundation
Grant awarded by the American Physical Society.}}
\medskip
\centerline{ Landau Institute for Theoretical Physics, Academy of Sciences, }
\centerline{ Kosygina 2, GSP-1, 117940 Moscow V-334, Russia
\footnote{$^{**}$}{E-mail: lashkevi@cpd.landau.free.msk.su}
}
\vskip 3 truecm
\par\noindent
Usual coset construction $\SU{k}\times\SU{l}/\SU{k+l}$ of Wess--Zumino
conformal field theory is presented as a coset construction of
minimal models. This new coset construction can be defined rigorously
and allows one to calculate easily correlation functions of a number of
primary fields.
\vfill\eject
\par\noindent
{\bf 1. Introduction}
\medskip
\pno
Coset construction\ref{1,2} is a method of obtaining a new
two-dimensional conformal field theory from two old ones, G and H.
Let $g$ and $h$ be chiral algebras of G and H models
respectively, $h$ be a subalgebra of $g$. States of
the coset theory are identified with orbits of the action of $h$
in representations of $g$. Conformal blocks\ref{3} of coset
construction can be obtained by some factorizing conformal blocks
of G theory\ref{4-7}.
\par
One of the most important examples of coset construction is\ref{1}
$$
\N{kl}\sim{\SU{k}\times\SU{l}\over\SU{k+l}},\eqno(1.1)
$$
where $\SU{k}$ means Wess--Zumino model\ref{8,9} with $\ $SU$(2)\ $ group
on level $k=1,2,3,\cdots$. Chiral algebra of Wess--Zumino $\SU{k}$
model coincides with $\sl{k}\simeq\hat{su}(2)_k^{\bf C}$ affine
Kac--Moody algebra
$$
[J_{k,m}^\aa,J_{k,n}^\bb]=f^{\aa\bb}_\ga J_{k,m+n}^\ga
+m{k\over2}g^{\aa\bb}\delta_{m+n,0}.
\eqno(1.2)
$$
Here $J_{k,n}^\aa$ are Laurent components of chiral currents,
$J_k^\aa(z)$,
$$
J_{k,m}^\aa=\oint{dz\over\2pi}z^mJ_k^\aa(z),\ \ \aa=+,-,0,
\eqno(1.3)
$$
with operator product expansion (OPE)
$$
J_k^\aa(z')J_k^\bb(z)={\textstyle{1\over2}kg^{\aa\bb}\over(z'-z)^2}
+{f^{\aa\bb}_\ga J_k^\ga(z)\over z'-z}+O(1);
\eqno(1.4)
$$
$f^{\aa\bb}_\ga$ are the structure constants and
$g^{\aa\bb}$ is the basic metric of $sl(2)$ algebra with nonzero components
$$
\matrix{
f_0^{+-}=2,\ f_+^{0+}=f_-^{-0}=1,\ g^{+-}=2,\ g^{00}=1,
\cr\cr
f^{\bb\aa}_\ga=-f^{\aa\bb}_\ga,\ g^{\bb\aa}=g^{\aa\bb}.
}\eqno(1.5)
$$
Chiral currents of the denominator of the coset construction (1.1)
are identified with diagonal currents of the numerator
$$
J_{k+l}^\aa(z)=J_k^\aa(z)+J_l^\aa(z).
\eqno(1.6)
$$
The central charge, $c$, of the Virasoro algebra
$$
[L_m,L_n]=(m-n)L_{m+n}+{c\over 12}m(m^2-1)\delta_{m+n,0}
\eqno(1.7)
$$
of N$_{kl}$ theory is given by
$$
c_{kl}=3-{6\over k+2}-{6\over l+2}+{6\over k+l+2}.
\eqno(1.8)
$$
Here $L_m$ are Laurent components of the energy-momentum tensor, $T(z)$,
$$
L_m=\oint{dz\over\2pi}z^{m+1}T(z),
\eqno(1.9)
$$
with OPE
$$
T(z')T(z)={\textstyle{1\over 2}c\over(z'-z)^4}
+{2T(z)\over(z'-z)^2}+{\partial T(z)\over z'-z}+O(1).
\eqno(1.10)
$$
\par
A special case of the coset construction (1.1) is
a unitary minimal conformal model
$$
\M{k}\sim\N{k1}\sim{\SU{k}\times\SU{1}\over\SU{k+1}}
\eqno(1.11)
$$
with central charge
$$
c_k=1-{6\over(k+2)(k+3)}.
\eqno(1.12)
$$
These models have been well investigated independently of their coset
structure by the use of bosonic representation.\ref{10-13} If $l=2$,  the
coset (1.1) yields $N=1$ superconformal models.\ref{14-16,2}
For general $l$ the coset construction $\N{kl}$ can be identified
with some bosonic models.\ref{17-21} Chiral algebras of these models
have not yet been found. These models will be referred to as
minimal-like ones. In principle, they are well defined by
their coset structure, but neither coset construction nor bosonization
gives an effective tool of calculating correlation functions.
Can we use the information about minimal models to calculate
correlation functions of $\N{kl}$ minimal-like models?
Let us make a naive transformation of the coset construction (1.1)
for $l=2$:\ref{22}
$$
\eqalign{
\N{k2}&\sim{\SU{k}\times\SU{2}\over\SU{k+2}}\times
{\SU{1}\times\SU{1}\times\SU{k+1}\over
\SU{1}\times\SU{1}\times\SU{k+1}}
\cr
&\sim{\SU{k}\times\SU{1}\over\SU{k+1}}
\times{\SU{k+1}\times\SU{1}\over\SU{k+2}}
\Bigg/{\SU{1}\times\SU{1}\over\SU{2}}
\cr
&\sim{\M{k}\M{k+1}\over\M{1}}.
}\eqno(1.13)
$$
In general case
$$
\N{kl}\sim{\M{k}\M{k+1}\cdots\M{k+l-1}
\over\M{1}\M{2}\cdots\M{l-1}}
\eqno(1.14)
$$
(the reason of absence of crosses will become clear later).
Thus far it has been nothing more than a trick with symbols.
There are two natural questions. Can we assign any exact meaning
to the ``fraction'' (1.14)? If it is possible, does the usual
coset construction yield the same result as (1.14)? In what follows
we answer in the affirmative to both questions. The main mathematical
trick is argued in Sec. 2, and it is applied to the coset construction
(1.14) in Sec. 3. Field identification of the coset construction
(1.14) and minimal-like models is done in Sec. 4. Sec. 5 answers
in the affirmative to the question of using information about
minimal models
for calculating correlation functions of minimal-like models.
\medskip
\pno
{\bf 2. Convolution and Factorization of Conformal Blocks}
\medskip
\pno
2.1. $Pairing\ Holomorphic\ and\ Antiholomorphic\ Parts$
\medskip
\pno
To make what follows more clear and to fix notations we shall recall
now some well known facts concerning connection between conformal
blocks and correlation functions. Consider conformal block
$F_{\{\la\mu\}}(\la_5;z)=F_{\la_1\mu_1,\la_2\mu_2}
^{\la_3\mu_3,\la_4\mu_4}(\la_5;z)$, $\{\la\mu\}=(\la_1\mu_1,
\la_2\mu_2,\la_3\mu_3,\la_4\mu_4)$ (see Fig. 1) of four fields
$\phi_{\la_1\mu_1}(0)$, $\phi_{\la_2\mu_2}(z)$, $\phi_{\la_3\mu_3}(1)$,
and $\phi_{\la_4\mu_4}(\infty)$ through states in $s$-channel from module
${\cal H}_{\la_5}$ with highest weight $\la_5$. Index $\mu_i$ labels states
in module ${\cal H}_{\la_i}$; $\mu_i$ will be usually omitted later.
Conformal blocks are solutions of differential equations imposed by
null-vectors of the chiral algebra.\ref{3,7,23} Conformal blocks
with different $\la_5$ permitted by fusion rules
form a basis of solution space of the differential equations.
Correlation function can
be found as\ref{3,10,11}\footnote{$^a$}{We only consider
primary fields of spin 0 for simplicity.
Furthermore, we consider each module to
coincide with the conjugate module and multiplicities to be
equal to 1. It is right for minimal, minimal-like and $SU(2)$
Wess--Zumino models. Generalizations are straitforward.}
$$
\eqalign{
\langle\phi_{\la_1}(0,0)
&\phi_{\la_2}(z,\overline{z})
\phi_{\la_3}(1,1)\phi_{\la_4}(\infty,\infty)\rangle
\cr
&=\sum_{\la_5}X_{\la_1\la_2}^{\la_3\la_4}(\la_5)
F_{\la_1\la_2}^{\la_3\la_4}(\la_5;z)
\overline{F_{\la_1\la_2}^{\la_3\la_4}(\la_5;z)},
}\eqno(2.1.1)
$$
where $X_{\la_1\la_2}^{\la_3\la_4}(\la_5)$ are coefficients,
and bar means complex conjugation. There are several equivalent
approaches to determing $X$-coefficients.
\par
$\underline{\hbox{1. Duality.}}$\ref{3} We can consider dual
($u$-channel) basis in solution space of differential equations,
$\left\{F_{\{\tilde{\la}\}}(\la_5;1-z)=
F_{\la_3\la_2}^{\la_1\la_4}(\la_5;1-z)\right\}$,
$\{\tilde{\la}\}=(\la_3,\la_2,\la_1,\la_4)$. Solutions of the
old basis can be expressed in terms of solutions of the new one (Fig. 2)
$$
F_{\la_1\la_2}^{\la_3\la_4}(\la_5;z)
=\sum_{\la_6}\aa_{\la_1\la_2}^{\la_3\la_4}(\la_5,\la_6)
F_{\la_3\la_2}^{\la_1\la_4}(\la_6;1-z).
\eqno(2.1.2)
$$
The coefficients $\aa_{\la_1\la_2}^{\la_3\la_4}(\la_5,\la_6)$
form so called braiding matrix $\aa_{\la_1\la_2}^{\la_3\la_4}$.
Substituting Eq. (2.1.2) into Eq. (2.1.1) and demanding that the
correlation function should be expressed in the same form by dual
blocks (Fig. 3), we obtain Dotsenko--Fateev equation\ref{10,11,24}
$$
\sum_{\la_5}X_{\la_1\la_2}^{\la_3\la_4}(\la_5)
\aa_{\la_1\la_2}^{\la_3\la_4}(\la_5,\la_6)
\overline{\aa_{\la_1\la_2}^{\la_3\la_4}(\la_5,\la'_6)}
=\delta_{\la_6\la'_6}X_{\la_3\la_2}^{\la_1\la_4}(\la_6).
\eqno(2.1.3)
$$
We shall cosider states from modules
${\cal H}_\la\otimes\overline{{\cal H}_\la}$
with the same $\la$ in holomorphic (${\cal H}$) and antiholomorphic
($\overline{{\cal H}}$) parts to be physical. We can say that all
intermediate states in $s$-channel in Eq. (2.1.1) are physical.
Eq. (2.1.3) ensures that only physical states should appear in $u$-channel.
In other words Eq. (2.1.3) ensures that there is no nonphysical
fields in OPE's of two physical fields.
\par
The normalization of four-point correlation functions is given by
the requirement
$$
\eqalign{
\langle\phi_{\la_1}(0,0)
&\phi_{\la_2}(z,\overline{z})
\phi_{\la_3}(1,1)\phi_{\la_4}(\infty,\infty)\rangle
\cr
&\approx\langle\phi_{\la_1}(0,0)\phi_{\la_2}(z,\overline{z})
\phi_{\la_m}(\infty,\infty)\rangle
\langle\phi_{\la_m}(0,0)
\phi_{\la_3}(1,1)\phi_{\la_4}(\infty,\infty)\rangle
\cr
&\hbox{if}\quad |z|\ll 1,
}\eqno(2.1.4)
$$
where $\phi_{\la_m}$ corresponds to the permitted intermediate state
with the lowest conformal dimension.
\par
If conformal blocks are normalized as
$$
F_{\{\la\}}(\la_5;z)\approx z^\delta,\ z\ll 1,
\eqno(2.1.5)
$$
the $X$-coefficients can be decomposed in product\ref{3,24}
$$
X_{\la_1\la_2}^{\la_3\la_4}(\la_5)=C_{\la_1\la_2\la_5}
C_{\la_5\la_3\la_4},
\eqno(2.1.6)
$$
where $C_{\la_1\la_2\la_3}$ are structure constants of the
operator algebra. Structure constants determine three-point correlation
functions
$$
\eqalign{
\langle\phi_{\la_1}(z_1,\overline{z}_1)
&\phi_{\la_2}(z_2,\overline{z}_2)
\phi_{\la_3}(z_3,\overline{z}_3)\rangle
\cr
&=C_{\la_1\la_2\la_3}(z_1-z_2)^{\De_3-\De_1-\De_2}
(z_1-z_3)^{\De_2-\De_1-\De_3}(z_2-z_3)^{\De_1-\De_2-\De_3}
\cr
&\times(\overline{z}_1-\overline{z}_2)
^{\overline{\De}_3-\overline{\De}_1-\overline{\De}_2}
(\overline{z}_1-\overline{z}_3)
^{\overline{\De}_2-\overline{\De}_1-\overline{\De}_3}
(\overline{z}_2-\overline{z}_3)
^{\overline{\De}_1-\overline{\De}_2-\overline{\De}_3},
}\eqno(2.1.7)
$$
where $\De_i$ and $\overline{\De}_i$ are holomorphic and
antiholomorphic conformal dimensions of the state $|\la_i\mu_i
\overline{\mu}_i\rangle$.
\par
$\underline{\hbox{2. Monodromy invariance.}}$\ref{10,11} A conformal
block can be decomposed in powers of $z$ for $|z|<1$
$$
F_{\{\la\}}(\la_5;z)=z^{\De_5-\De_1-\De_2}
\sum_{n=0}^{\infty}\bb_nz^n.
\eqno(2.1.8)
$$
Therefore
$$
F_{\{\la\}}(\la_5;e^{\2pi}z)=e^{\2pi(\De_5-\De_1-\De_2)}
F_{\{\la\}}(\la_5;z).
\eqno(2.1.9)
$$
Thus the monodromy invariance of the correlation function (2.1.1)
under moving $z$ round zero is evident. To investigate monodromy
properties under moving $z$ round one we can substitute Eq. (2.1.2)
into Eq. (2.1.1). The requirement of monodromy invariance yields
again Dotsenko--Fateev equation (2.1.3).
\par
$\underline{\hbox{3. Vertex operators, quantum group etc.}}$\ref{13,24-27}
Instead of conformal blocks vertex operators,
$$
\left(\phi_{\la\mu}(z)\right)_{\la'}^{\la''}:{\cal H}_{\la'}
\longrightarrow{\cal H}_{\la''},
\eqno(2.1.10)
$$
can be considered. The primary field, $\phi_\la(z,\overline{z})$,
is expressed in terms of vertex operators as
$$
\phi_\la(z,\overline{z})=\bigoplus_{\la'\la''}
X_\la(\la'',\la')\left(\phi_{\la}(z)\right)_{\la'}^{\la''}
\overline{\left(\phi_{\la}(z)\right)_{\la'}^{\la''}}.
\eqno(2.1.11)
$$
The direct sum means that each term acts in its own pair of modules.
Four-point conformal blocks are given by
$$
F_{\{\la\}}(\la_5;z)=\Big\langle
\left(\phi_{\la_4}(\infty)\right)_{\la_4}^{0}
\left(\phi_{\la_3}(1)\right)_{\la_5}^{\la_4}
\left(\phi_{\la_2}(z)\right)_{\la_1}^{\la_5}
\left(\phi_{\la_1}(0)\right)_{0}^{\la_1}
\Big\rangle,
\eqno(2.1.12)
$$
index 0 means the vacuum module. Transposing vertex operators in
a conformal block we change basis in the solution space of the
differential equations. Therefore
$$
\left(\phi_{\la_3}(z')\right)_{\la_5}^{\la_4}
\left(\phi_{\la_1}(z)\right)_{\la_2}^{\la_5}
=\sum_{\la_6}R_{\la_1\la_2}^{\la_3\la_4}(\la_5,\la_6)
\left(\phi_{\la_1}(z)\right)_{\la_6}^{\la_4}
\left(\phi_{\la_3}(z')\right)_{\la_2}^{\la_6}.
\eqno(2.1.13)
$$
It is easy to prove that
$$
\aa_{\la_1\la_2}^{\la_3\la_4}(\la_5,\la_6)
=R_{\la_10}^{\la_2\la_5}(\la_1,\la_2)
R_{\la_1\la_2}^{\la_3\la_4}(\la_5,\la_6)
R_{\la_20}^{\la_3\la_6}(\la_2,\la_3).
\eqno(2.1.14)
$$
Braiding matrix and R-matrix coincide up to a normalization factor.
$X$-factors for conformal blocks are given by
$$
X_{\la_1\la_2}^{\la_3\la_4}(\la_5)
=X_{\la_4}(0,\la_4)X_{\la_3}(\la_4,\la_5)X_{\la_2}(\la_5,\la_1)
X_{\la_1}(\la_1,0).
\eqno(2.1.15)
$$
If confomal blocks are normalized as in Eq. (2.1.5), then
$$
X_{\la_1}(\la_3,\la_2)=C_{\la_1\la_2\la_3},
\quad X_\la(\la,0)=X_\la(0,\la)=1.
\eqno(2.1.16)
$$
It can be proved\ref{25-27} that R-matrices of $\SU{k}$ model
coincide with those of representations of $\slq{q(k)}$ quantum
group (more precisely, quantum enveloping algebra) with
$$
q(k)=\exp\left({\2pi\over k+2}\right),
\eqno(2.1.17)
$$
and R-matrices of the minimal model $\M{k}$ coincide with those
of representations of $\slq{q(k)}\times\slq{\overline{q(k+1)}}$
quantum group (see also Appendix A). Generators of quantum
group act in the spaces spanned by vertex operators.
\medskip
\pno
2.2. $Convolution\ of\ Conformal\ Blocks\ and\ Vertex\ Operators$
\medskip
\pno
Consider two conformal theories $\Psi$ and $\Phi$ with conformal
blocks $\Psi_{\{l\la,m\}}(l_5\la_5;z)$ and
$\Phi_{\{\la L,M\}}(\la_5L_5;z)$ respectively. Indices $m$ and
$M$ label vectors in spaces respectively ${\cal H}_{l\la}^\Psi$ and
${\cal H}_{\la L}^\Phi$, each module is labelled by two indices.
Suppose braiding matrices $\aa$ of $\Psi$ theory and
$\bb$ of $\Phi$ theory to be decomposed as
$$
\eqalignno{
\aa_{\{l\la\}}(l_5\la_5,l_6\la_6)
&=A_{\{l\}}(l_5,l_6)K_{\{\la\}}(\la_5,\la_6),
&(2.2.1)
\cr
\bb_{\{\la L\}}(\la_5L_5,\la_6L_6)
&=\overline{K_{\{\la\}}(\la_5,\la_6)}B_{\{L\}}(L_5,L_6).
&(2.2.2)}
$$
In the language of quantum groups it means that the quantum group
of the theory $\Psi$ is the direct product $U_{q_1}\times W_{q_3}$
with deformation parameters $q_1$ and $q_3$, and the quantum
group of the theory $\Phi$ is $W_{\overline{q_3}}\times V_{q_2}$.
If we forgot for a while about $A$ and $B$ ($U$ and $V$) factors
we would see that two holomorphic conformal blocks behave just
as holomorphic and antiholomorphic parts of correlation function.
More precisely, consider the solution, $Z_{\{\la\}}(\la_5)$,
of the Dotsenko--Fateev equation for $K$-factor
$$
\sum_{\la_5}Z_{\la_1\la_2}^{\la_3\la_4}(\la_5)
K_{\la_1\la_2}^{\la_3\la_4}(\la_5;\la_6)
\overline{K_{\la_1\la_2}^{\la_3\la_4}(\la_5;\la'_6)}
=\delta_{\la_6\la'_6}Z_{\la_3\la_2}^{\la_1\la_4}(\la_6).
\eqno(2.2.3)
$$
We construct new conformal blocks\ref{5-7,21,22}
$$
F_{\{l\la L\}}(l_5L_5;z)=\sum_{\la_5}Z_{\{\la\}}(\la_5)
\Psi_{\{l\la\}}(l_5\la_5;z)\Phi_{\{\la L\}}(\la_5L_5;z).
\eqno(2.2.4)
$$
Note that $\la_5$ is absent in the l.h.s. The braiding matrix, $\ga$,
of new conformal blocks is given by
$$
\ga_{\{l L\}}(l_5L_5,l_6L_6)=A_{\{l\}}(l_5,l_6)B_{\{L\}}(L_5,L_6).
\eqno(2.2.5)
$$
Respective quantum group is $U_{q_1}\times V_{q_2}$.
\par
In terms of vertex operators we construct new vertices $f_{l\la L}(z)$
from old ones $\psi_{l\la}(z)$ and $\phi_{\la L}(z)$ as
$$
\left(f_{l\la L}(z)\right)^{l''L''}_{l'L'}
=\bigoplus_{\la'\la''}Z_\la(\la'',\la')
\left(\psi_{l\la}(z)\right)^{l''\la''}_{l'\la'}
\left(\phi_{\la L}(z)\right)^{\la''L''}_{\la'L'},
\eqno(2.2.6)
$$
where $Z_{\{\la\}}(\la_5)=
Z_{\la_4}(0,\la_4)Z_{\la_3}(\la_4,\la_5)Z_{\la_2}(\la_5,\la_1)
Z_{\la_1}(\la_1,0)$. We shall designate such ``convolution''
of vertices as
$$
f_{l\la L}(z)=\psi_{l\la}(z)\phi_{\la L}(z).
\eqno(2.2.7)
$$
It will not cause confusion, because the usual tensor
product of vertices will not occur in this paper.
\par
Note that Eq. (2.1.11) can be written in these designations as
$$
\phi_\la(z,\overline{z})=\phi_\la(z)\overline{\phi_\la(z)}.
\eqno(2.2.8)
$$
At last, there is a relationship between structure constants
$$
C^F_{l_1\la_1L_1,l_2\la_2L_2,l_3\la_3L_3}
=C^\Psi_{l_1\la_1,l_2\la_2,l_3\la_3}
C^\Phi_{\la_1L_1,\la_2L_2,\la_3L_3}
\eqno(2.2.9)
$$
of theories $F$, $\Psi$ and $\Phi$ respectively.
\par
Consider examples. Quantum groups of models $\M{k}$ and $\SU{k+1}$
are \pno $\slq{q(k)}\times\slq{\overline{q(k+1)}}$ and $\slq{q(k+1)}$
respectively. Their convolution gives vertex operators of
a theory with the quantum group $\slq{q(k)}$ or, taking into
account that $\slq{q(1)}$ corresponds to trivial monodromy (or braiding)
properties (Appendix B), with the quantum group
$\slq{q(k)}\times\slq{q(1)}$. But it is just the quantum group of
the numerator of the coset construction (1.11).
\par
The second example is convolution of $\M{k}$ and $\M{k+1}$
models with quantum qroups $\slq{q(k)}\times\slq{\overline{q(k+1)}}$
and $\slq{q(k+1)}\times\slq{\overline{q(k+2)}}$ respectively.
This convolution is the most important in this work.
\medskip
\pno
2.3. $Factorization$
\medskip
\pno
Now we shall consider the reverse to convolution procedure
of factorizing conformal blocks. Suppose that $F$ and $\Phi$
conformal blocks from previous subsection are known, and (2.2.4)
is considered as an equation in unknown $\Psi$.\ref{5-7}
Can it be solved? How many solutions does it have? It turns
out that there is a unique solution if chiral currents of
$\Phi$ theory form a subalgebra of chiral algebra of $F$
theory, and if Virasoro generators, $L^\Phi_m$, from this
subalgebra commute with $L^\Psi_n=L^F_n-L^\Phi_n$, where
$L^F_n$ are Virasoro generators of the theory $F$.\ref{7}
This is the coset construction condition.
Now we shall sketch the proof. To avoid too
cumbersome designations we shall omit indices $\la_1,\cdots,
\la_4$ and all $l_i$, $L_i$ and keep indices $M_i$ labelling
basic vectors in the space ${\cal H}_{\la_i L_i}^\Phi$.
Eq. (2.2.4) is written as
$$
F_{\{M\}}(z)=\sum_{\la_5}Z(\la_5)\Psi(\la_5;z)
\Phi_{\{M\}}(\la_5;z).
\eqno(2.3.1)
$$
Let us introduce matrix designations $F_i(z)=F_{\{M\}}(z)$,
$i=\{M\}$; $\Psi_i(z)=\Psi(\la_5;z)$, $i=\la_5$;
$\Phi_{ij}(z)=\Phi_{\{M\}}(\la_5;z)$, $i=\{M\}$, $j=\la_5$;
$Z_{ij}=\delta_{\la_5\la'_5}Z(\la_5)$, $i=\la_5$, $j=\la'_5$.
We see that
$$
F=\Phi Z\Psi.
\eqno(2.3.2)
$$
Now we must restrict region of varying of index $\{M\}$ to a finite
number of values which is equal to the number of values of $\la_5$
permitted by fusion rules. The restriction must allow
other rows of $\Phi$ or $F$ to be restored
using Ward identities\ref{3,9} for
chiral currents of $\Phi$ theory. Columns of restricted matrix
$\Phi$ are linearly independent solutions of the differential
equations. Hence its determinant is a Wronskian of linearly
independent solutions, and
$$
\det\Phi^{{\rm restr.}}\neq 0,\hbox{ if }z\neq 0,1,\infty.
\eqno(2.3.3)
$$
By definition $\ \det Z\neq 0$, and we obtain
$$
\Psi=Z^{-1}\Phi^{-1}F
\eqno(2.3.4)
$$
for restricted matrices. This proves uniqueness.
\par
Every chiral current of $\Phi$ theory can be presented as differential
operator which acts on a conformal block. Let us replace there each
derivative by the operator $[L^\Phi_{-1},\dot]$. We obtain the same
relationships between rows of both full matrices $F$ and $\Phi$.
It assures $\Psi$ from Eq. (2.3.4) to be independent of restriction.
It means that Eq.  (2.3.1) has a solution.
\medskip
\pno
{\bf 3. Minimal-Like Models as Coset Constructions of Minimal Models}
\medskip
\pno
The minimal model $\M{k}$ is described by vertex operators
$$
\eqalign{
&\left(\phi^{(k)}_{(p,q)}(z)\right)_{m,n}:{\cal H}_{(p_1,q_1)}
\longrightarrow{\cal H}_{(p_1+p-1-2m,q_1+q-1-2n)},
\cr
&p=1,2,\cdots,k+1;\ q=1,2,\cdots,k+2;
\cr
&m=0,1,\cdots,p-1;
\ n=0,1,\cdots,q-1
}\eqno(3.1)
$$
with conformal dimensions
$$
\De_{(p,q)}={[(k+3)p-(k+2)q]^2-1\over 4(k+2)(k+3)}.
\eqno(3.2)
$$
There is an equivalence
$$
\eqalign{
{\cal H}_{(k+2-p,k+3-q)}
&\sim{\cal H}_{(p,q)},
\cr
\left(\phi^{(k)}_{(k+2-p,k+3-q)}(z)\right)_{p_1-1-m,q_1-1-n}
\Big|_{{\cal H}_{(p_1,q_1)}}
&\sim\left(\phi^{(k)}_{(p,q)}(z)\right)_{m,n}
\Big|_{{\cal H}_{(p_1,q_1)}}.
}\eqno(3.3)
$$
It means that any state corresponds to two vectors:
one from ${\cal H}_{(p,q)}$ and one from
${\cal H}_{(k+2-p,k+3-q)}$.
\par
Variables $p$, $q$, $m$, $n$ possess quantum group sense.
Namely, ``momenta'' (components of highest weight) $J_1$, $J_2$ and
``projections of momenta'' (components of weight) $M_1$, $M_2$ for
an irreducible representation of $\slq{q(k)}\times\slq{\overline{q(k+1)}}$
realized by vertex operators for given $(p,q)$ are given by
$$
\eqalign{
&J_1=\textstyle{1\over 2}(p-1),\ M_1=J_1-m,
\cr
&J_2=\textstyle{1\over 2}(q-1),\ M_2=J_2-m.
}\eqno(3.4)
$$
\par
The minimal-like model $\N{kl}$ is described by vertices
$$
\eqalign{
&\left(\phi^{(k,l)}_{pp'q}(z)\right)_{mm'n}:{\cal H}_{p_1p'_1q_1}
\longrightarrow{\cal H}_{p_1+p-1-2m,p'_1+p'-1-2m',q_1+q-1-2n},
\cr
&p=1,2,\cdots,k+1;\ p'=1,2,\cdots,l+1;\ q=1,2,\cdots,k+l+1;
\cr
&p+p'-q-1\in 2{\bf Z};
\cr
&m=0,1,\cdots,p-1;
\ m'=0,1,\cdots,p'-1;
\ n=0,1,\cdots,q-1.
}\eqno(3.5)
$$
There is an equivalence
$$
\eqalign{
&{\cal H}_{k+2-p,l+2-p',k+l+2-q}
\sim{\cal H}_{pp'q},
\cr
&\eqalign{
\left(\phi^{(k,l)}_{k+2-p,l+2-p',k+l+2-q}(z)
\right)
&_{p_1-1-m,p'_1-1-m',q_1-1-n}
\Big|_{{\cal H}_{p_1p'_1q_1}}
\cr
&\sim\left(\phi^{(k,l)}_{pp'q}(z)\right)_{mm'n}
\Big|_{{\cal H}_{p_1p'_1q_1}}.
}}\eqno(3.6)
$$
If $l=1$, the minimal-like theory $\N{kl}$ reduces to $\M{k}$:
$$
\matrix{
\left(\phi^{(k)}_{(p,q)}(z)\right)_{m,n}
\sim\displaystyle{\sum_{m'}}
a_{m'}\left(\phi^{(k,l)}_{pp'q}(z)\right)_{mm'n},
\cr\cr
p'-1=p-q(\hbox{mod }2).
}\eqno(3.7)
$$
Bosonic representations for minimal and minimal-like theories,
conformal dimensions of $\phi_{pp'q}(z)$ and rules for calculating
three-point correlation functions are presented in Appendix C.
\par
It has been mentioned in Sec. 2.2 that there exists a
convolution $\M{k}\M{k+1}$ of two minimal models. It consists
of vertex operators $\phi^{(k)}_{(p,s)}(z)\phi^{(k+1)}_{(s,q)}(z)$.
We can consider a chain of such convolutions
$\M{k}\M{k+1}\cdots\M{k+l-1}$ with vertex operators
$$
\phi_{(p,s,q)}(z)=\phi^{(k)}_{(p,s_1)}(z)\phi^{(k+1)}_{(s_1,s_2)}(z)
\cdots\phi^{(k+l-1)}_{(s_{l-1},q)}(z),\ s=(s_1,\cdots,s_{l-1}).
\eqno(3.8)
$$
This is the numerator of the coset construction (1.14). Its
quantum group is $\slq{q(k)}\times\slq{\overline{q(k+l)}}$.
This quantum group differs from those of $\N{kl}$ by absence
of $\slq{q(l)}$ factor. We must ``glue it up'' in $\N{kl}$
by the denominator of the coset construction. Namely, consider
vertex operators
$$
\psi_{t;pq}(z)=\phi^{(1)}_{(t_1,t_2)}(z)\phi^{(2)}_{(t_2,t_3)}(z)
\cdots\phi^{(l-1)}_{(t_{l-1},p')}(z)\phi^{(k,l)}_{pp'q}(z),
\ t=(t_1,\cdots,t_{l-1},t_l=p').
\eqno(3.9)
$$
We want to prove that vertices $\psi_{t;pq}(z)$ realize
some subtheory $\M{1}\M{2}\cdots\M{l-1}\N{kl}$ of the
theory $\M{k}\M{k+1}\cdots\M{k+l-1}$. To do it we shall
find the energy-momentum tensors of $\M{k}$, $\M{k+1}$,
$\dots$, $\M{k+l-1}$ models among fields of the theory
$\M{1}\M{2}\cdots\M{l-1}$ $\cdot\N{kl}$. Note that there is a lot
of chiral currents in this theory. Particularly, every field
$$
J_t(z)=\psi_{t;11}(z)
\eqno(3.10)
$$
can be considered as a chiral current. Indeed, $J_T(z)$ realizes
the unit representation of $\slq{q(1)}\times\slq{\overline{q(l)}}$
quantum group. In other words, expansion of the operator product
$J_T(z')\psi_{t;pq}(z)$ gives fields with conformal
dimensions which differ from conformal dimension of
$\psi_{t;pq}(z)$ by an integer. This can be easily seen using
the formulae for conformal dimensions from Appendix C and
standard fusion rules produced by quantum group $\slq{q}$.
Therefore, if $z'$ goes round $z$ and returns to the initial
point, a conformal block which contains this product does not
change. It means that $J_t(z)$ has no need of pairing with an
antiholomorphic field and it can be included into chiral algebra.
\par
We shall use induction. Eq. (1.14) can be rewritten as
$$
\N{kl}\sim{\N{k,l-1}\M{k+l-1}\over\M{l-1}}.
\eqno(3.11)
$$
We shall not prove this equation thoroughly at each step
of induction, but
we shall construct energy-momentum tensors of  $\N{k,l-1}$ and
$\M{k+l-1}$ from fields of $\N{kl}$ and $\M{l-1}$ and prove
OPE's of fields necessary for the following step.
\par
Consider the current of type (3.10)
$$
J^{(k,l)}_3(z)\sim\phi^{(l-1)}_{(1,3)}(z)\phi^{(k,l)}_{131}(z)
\eqno(3.12)
$$
with normalization
$$
\Big\langle J^{(k,l)}_3(z')J^{(k,l)}_3(z)\Big\rangle
=(z'-z)^{-4}.
\eqno(3.13)
$$
It can be proved using formulae from Appendix C that
$$
\eqalign{
J^{(k,l)}_3(z')J^{(k,l)}_3(z)
&={1\over(z'-z)^4}+{2\theta_3(z)\over(z'-z)^2}
+{\partial\theta_3(z)\over z'-z}+O(1),
\cr
\theta_3(z)
&={l(l+1)\over(l-1)(l+4)}T_{l-1}(z)
+{(k+2)(l+4)(k+l+2)\over 3kl(k+l+4)}T_{k,l}(z)
\cr
&+{2(2k+l+4)(l-2)\over\sqrt{3kl(l-1)(l+4)(k+l+4)}}J^{(k,l)}_3(z),
}\eqno(3.14)
$$
where $T_k(z)$ and $T_{k,l}(z)$ are energy-momentum tensors
of $\M{k}$ and $\N{kl}$ models respectively. Using Eqs.
(1.10), (1.12), (1.8) and (3.14) we obtain that fields
$$
\eqalign{
T_{k+l-1}(z)
&={(l+1)(k+l+4)\over(l+4)(k+l+1)}T_{l-1}(z)
+{k+2\over l(k+l+1)}T_{k,l}(z)
\cr
&+{1\over k+l+1}\sqrt{3{k(l-1)(k+l+4)\over l(l+4)}}\ J^{(k,l)}_3(z),
\cr
T_{k,l-1}(z)
&={3k\over(l+4)(k+l+1)}T_{l-1}(z)
+{(l-1)(k+l+2)\over l(k+l+1)}T_{k,l}(z)
\cr
&-{1\over k+l+1}\sqrt{3{k(l-1)(k+l+4)\over l(l+4)}}\ J^{(k,l)}_3(z)
}\eqno(3.15)
$$
obey Eq. (1.10) for $\M{k+l-1}$ and $\N{k,l-1}$ models, and besides
$$
T_{k+l-1}(z')T_{k,l-1}(z)=O(1).
\eqno(3.16)
$$
We must now prove that the field $\phi^{(k,l-1)}_{131}(z)$
necessary to construct $J^{(k,l-1)}_3$ for the
next step of induction can be expressed
in terms of $\phi^{(k,l)}_{131}(z)$.
It is enough for our purposes to restore coefficients at
$\phi^{(k,l-1)}_{111}(z)=1$ and $\phi^{(k,l-1)}_{131}(z)$ in
expansion of $\phi^{(k,l-1)}_{131}(z')\phi^{(k,l-1)}_{131}(z)$
operator product [see (C.26)]. In the normalization of the
bosonic representation we reach it by identification
$$
\phi^{(k,l-1)}_{131}(z)={k\over l+4}\phi^{(l-1)}_{(3,1)}(z)
\phi^{(k,l)}_{111}(z)
+{1\over l+2}\phi^{(l-1)}_{(3,3)}(z)\phi^{(k,l)}_{131}(z).
\eqno(3.17)
$$
Now we can, in principle, express all energy-momentum tensors
of $\M{k}$, $\M{k+1}$, $\dots$, $\M{k+l-1}$ in terms of
chiral currents of $\M{1}\M{2}\cdots\M{l-1}\N{kl}$.
This completes the proof.
\par
Now we shall see that the energy-momentum tensors of the
denominator of the coset construction (1.14) and of $\N{kl}$
can be expressed in terms of chiral currents of the numerator.\footnote
{$^b$}{Recall that minimal-like models with $k,l>1$ are not
minimal ones and are not defined uniquely by Virasoro algebra.}
Namely, we have
$$
\eqalign{
T_{l-1}(z)
&={(l-1)(k+l+2)\over(k+2)(k+l-1)}T_{k+l-1}(z)
+{k+2\over(l+2)(k+l+3)}T_{k,l-1}(z)
\cr
&+{1\over l+2}\sqrt{3{k(l-1)(k+l+4)\over(k+l-1)(k+l+3)}}\ t_{k,l-1}(z),
\cr
T_{k,l}(z)
&={3k\over(k+2)(k+l-1)}T_{k+l-1}(z)
+{(l+1)(k+l+4)\over(l+2)(k+l+3)}T_{k,l-1}(z)
\cr
&-{1\over l+2}\sqrt{3{k(l-1)(k+l+4)\over(k+l-1)(k+l+3)}}\ t_{k,l-1}(z),
}\eqno(3.18)
$$
where
$$
t_{k,l}(z)\sim\phi^{(k,l)}_{113}(z)\phi^{(k+l)}_{(3,1)}(z)
\eqno(3.19)
$$
is a chiral current, and
$$
\eqalign{
t_{k,l}(z')t_{k,l}(z)
&={1\over(z'-z)^4}+{2\theta(z)\over(z'-z)^2}
+{\partial\theta(z)\over z'-z}+O(1),
\cr
\theta(z)
&={(k+l+3)(k+l+4)\over(k+l)(k+l+5)}T_{k+l}(z)
+{(k+2)(l+2)(k+l)\over 3kl(k+l+4)}T_{k,l}(z)
\cr
&+{2(k-l)(k+l+6)\over\sqrt{3kl(k+l)(k+l+4)(k+l+5)}}t_{k,l}(z).
}\eqno(3.20)
$$
Let us introduce more general currents
$$
\matrix{
t_{k,l}^{(n)}\sim\phi^{(k,l)}_{113}(z)
\displaystyle{\prod_{i=0}^{n-1}}\phi^{(k+l+i)}_{(3,3)}(z)
\cdot\phi^{(k+l+n)}_{(3,1)}(z),
\cr\cr
\langle t_{k,l}(z')t_{k,l}(z)\rangle=(z'-z)^{-4},
}\eqno(3.21)
$$
such that the proportionality coefficient should be negative in bosonic
representation. In the same way as Eq. (3.17)
we obtain
$$
t_{k,l}^{(n)}(z)=\sqrt{k\over l(k+l-1)}\ t_{k+l-n-1}^{(n)}(z)
-\sqrt{(l-1)(k+l)\over l(k+l-1)}\ t_{k,l-1}^{(n+1)}(z),
\eqno(3.22)
$$
where
$$
t_k^{(n)}=t_{k,1}^{(n)}
\sim\phi^{(k)}_{(1,3)}(z)\prod_{i=1}^n\phi^{(k+i)}_{(3,3)}(z)
\cdot\phi^{(k+n+1)}_{(3,1)}(z).
\eqno(3.23)
$$
Applying Eqs. (3.18), (3.22) repeatedly we have
$$
\eqalign{
T_{k,l}(z)
&={k(k+l+4)\over l+2}\Bigg(\sum_{m=0}^{l-1}
{3T_{k+m}(z)\over(k+m)(k+m+5)}
\cr
&-\sum_{m=1}^{l-1}\sqrt{3\over(k+m+4)(k+m+5)}
\sum_{n=0}^{m-1}{(-)^nt_{k+m-n-1}^{(n)}\over\sqrt{(k+m-n)(k+m-n-1)}}
\Bigg),
\cr\cr
T_l(z)
&=T_{k,l-1}(z)+T_{k+l-1}(z)-T_{k,l}(z).
}\eqno(3.24)
$$
\medskip
\pno
{\bf 4. Field Identification}
\medskip
\pno
Now we can identify some fields of minimal-like models with
coset fields. Let us try to find fields which are primary with
respect to $T_{k,l}(z)$, $T_{l-1}(z)$, $\dots$, $T_{1}(z)$
in the space spanned by fields $\phi_{(p,s,q)}(z)
=\phi^{(k)}_{(p,s_1)}(z)\cdots\phi^{(k+l-1)}_{(s_{l-1},q)}(z)$.
\par
Consider fusion rule
$$
T(z')\phi_{(p,s,q)}(z)\sim\sum_{s'(\forall i:\ s'_i-s_i=0,\pm 2)}
\left[\phi_{(p,s',q)}(z)\right],
\eqno(4.1)
$$
where $T(z)$ is one of the fields
$T_{k,l}(z)$, $T_{l-1}(z)$, $\dots$, $T_{1}(z)$; brackets mean
conformal family of the field. The absence of fields with
$s'_i-s_i\neq 0,\pm 2$ is related with absence of vertices
$\phi^{(k+i)}_{(p,q)}(z)$ with $p,q\neq 1,3$ in
currents $t_{k+j}^{(n)}$. Operators $L_m$, $m>0$ annihilate
$\phi_{(p,s,q)}(0)$ if
$$
\De_{(p,s,q)}\leq\De_{(p,s',q)},\ s'_i-s_i=0,\pm 2,
\eqno(4.2)
$$
where conformal dimensions $\De_{(p,s,q)}$ in
$\M{k}\M{k+1}\cdots\M{k+l-1}$ theory are given by
($s_0=p$, $s_l=q$)
$$
\eqalign{
\De_{(p,s,q)}
&=\sum_{i=0}^{l-1}\De_{(s_i,s_{i+1})}^{(k+i)}
\cr
&={1\over 4}{k+3\over k+2}p^2
+{1\over 4}{k+l+1\over k+l+2}q^2+{1\over 2}\sum_{i=1}^{l-1}s_i^2
-{1\over 2}ps_1-{1\over 2}s_{l-1}q
-{1\over 2}\sum_{i=1}^{l-2}s_is_{i+1}.
}\eqno(4.3)
$$
\par
Now we shall cosider special cases.
\par
$\underline{1.\ \De_{(p,s,q)}\leq \De_{(p,s',q)}-2,\ s'_i-s_i=0,\pm 2}.$
It is easy to prove that
$$
s_i=p+{q-p\over l}i,\ {q-p\over l}\in{\bf Z}.
\eqno(4.4)
$$
The conformal dimension of the respective primary field
calculated using Eq. (3.24) is given by
$$
\De_{pq}={[(k+l+2)p-(k+2)q]^2-l^2\over 4l(k+2)(k+l+2)}.
\eqno(4.5)
$$
Eq. (C.25) gives the same result:
$$
\De_{pq}=\De^{(k,l)}_{pp'q},
\quad p'-1=p-q(\hbox{mod }2l)=1\hbox{ or }l+1.
\eqno(4.6)
$$
Conformal dimension, $\De_{(p,s,q)}$, of the ``numerator field''
coincides with $\De^{(k,l)}_{pp'q}$, and hence
$$
\De^{(m)}_{(t_i,t_i+1)}=0,\quad m=1,2,\cdots,l-1.
\eqno(4.7)
$$
Finally we obtain
$$
\matrix{
\phi^{(k,l)}_{pp'q}(z)=\phi^{(k)}_{(p,s_1)}(z)
\phi^{(k+1)}_{(s_1,s_2)}(z)\cdots\phi^{(k+l-1)}_{(s_{l-1},q)}(z),
\cr\cr
s_i=p+\displaystyle{i\over l}(q-p),\ q-p\in l{\bf Z},
\ p'-1=p-q(\hbox{mod }2l).
}\eqno(4.8)
$$
\par
$\underline{2.\ \De_{(p,s,q)}\leq \De_{(p,s',q)}-1,\ s'_i-s_i=0,\pm 2}.$
In this case one can find representations of fields
$\phi^{(k,l)}_{pp'q}(z)$ with indices
$$
p'=r+1,\ r=p-q(\hbox{mod }2l)
\eqno(4.9{\rm a})
$$
or
$$
p'=l-r+1,\ l+r=p-q(\hbox{mod }2l),
\eqno(4.9{\rm b})
$$
and conformal dimensions
$$
\De^{(k,l)}_{pp'q}={[(k+l+2)p-(k+2)q]^2-l^2\over 4l(k+2)(k+l+2)}
+{r(l-r)\over 2l(l+2)}.
\eqno(4.10)
$$
Two simplest representations for these fields are
$$
\prod_{m=r}^{l-1}\phi^{(m)}_{(r+1,r+1)}(z)\cdot\phi^{(k,l)}_{pp'q}(z)
=\phi^{(k)}_{(p,s_1)}(z)\prod_{i=1}^{l-2}
\phi^{(k+i)}_{(s_i,s_{i+1})}(z)\cdot\phi^{(k+l-1)}_{(s_{l-1},q)}(z)
\eqno(4.11{\rm a})
$$
with
$$
s_i=\cases{
p+\left({q-p+r\over l}-1\right)i
& if $i\leq r$;
\cr
p+r+{q-p+r\over l}i
& if $i\geq r$,}
\eqno(4.11{\rm b})
$$
(Fig. 4a) and
$$
\prod_{m=l-r}^{l-1}\phi^{(m)}_{(l-r+1,l-r+1)}(z)
\cdot\phi^{(k,l)}_{pp'q}(z)
=\phi^{(k)}_{(p,s_1)}(z)\prod_{i=1}^{l-2}
\phi^{(k+i)}_{(s_i,s_{i+1})}(z)\cdot\phi^{(k+l-1)}_{(s_{l-1},q)}(z)
\eqno(4.12{\rm a})
$$
with
$$
s_i=\cases{
p+{q-p+r\over l}i
& if $i\leq l-r$;
\cr
p+l-r+\left({q-p+r\over l}-1\right)i
& if $i\geq l-r$,}
\eqno(4.12{\rm b})
$$
(Fig. 4b). If $r=0$ or $r=l$, we return to the case 1.
\par
Consider $i$-dependence of $s_i$ as in Fig. 5. The plot consists
of segments with slopes ${1\over l}(q-p+r)-1$  and ${1\over l}(q-p+r)$.
There is a representative for each such plot
$$
\eqalign{
&\prod_{m=1}^{l-1}\phi^{(m)}_{(t_m,t_m)}(z)
\cdot\phi^{(k,l)}_{pp'q}(z)
=\phi^{(k)}_{(p,s_1)}(z)\prod_{i=1}^{l-2}\phi^{(k+i)}_{(s_i,s_{i+1})}(z)
\cdot\phi^{(k+l-1)}_{(s_{l-1},q)}(z),
\cr
&t_m-1=s_m-p-\left({q-p+r\over l}-1\right)m
\quad \hbox{if }s_m-s_{m-1}={q-p+r\over l}-1,
\cr
&t_m-1=p-s_m-{q-p+r\over l}m\quad
\hbox{ if }s_m-s_{m-1}={q-p+r\over l}.
}\eqno(4.13)
$$
We let $t_m=1$ if $t_m>m$ according these formulae. It is easy
to prove that
$$
\eqalign{
t_{l-1}
&=p'\quad \hbox{or}\quad t_{l-1}=l+2-p',
\cr
t_{m-1}
&=t_m\quad \hbox{or}\quad t_{m-1}=m+2-t_m.
}\eqno(4.14)
$$
Taking into account the equivalence (3.3) we see that Eq. (4.14)
is consistent with Eq. (3.9).
\par
$\underline{3.\ \De_{(p,s,q)}\leq \De_{(p,s',q)},\ s'_i-s_i=0,\pm 2}.$
This case is very difficult for calculation and has not been investigated
thoroughly. We discuss it only for $l=2$.
\par
We must consider [in addition to fields (4.8), (4.11$-$13)]
fields \pno $\phi_{(p,s_\pm,q)}(z)$ with
$$
s_\pm=\textstyle{1\over 2}(p+q)\pm 1.
\eqno(4.15)
$$
We have
$$
\De_{(p,s_+,q)}=\De_{(p,s_-,q)},
\eqno(4.16)
$$
and we must look for eigenvectors of
$L_0^{(k,2)}=\oint{dz\over\2pi}T_{k,2}(z)$ in the space spanned
by $\phi_{(p,s_+,q)}(0)$ and $\phi_{(p,s_-,q)}(0)$. We find
$$
\matrix{
\phi^{(1)}_{(2,1)}(z)\phi^{(k,2)}_{pp'q}(z)
=\sqrt{\textstyle{1\over 2}+y}\ \phi_{(p,s_+,q)}(z)
-\sqrt{\textstyle{1\over 2}-y}\ \phi_{(p,s_-,q)}(z),
\cr\cr
y=[(k+4)p-(k+2)q]^{-1}, p'-1=p-q(\hbox{mod }4),
\ p-q\in 2{\bf Z},
}\eqno(4.17)
$$
and
$$
\matrix{
\phi^{(k,2)}_{pp'q}(z)
=\sqrt{\textstyle{1\over 2}-y}\ \phi_{(p,s_+,q)}(z)
+\sqrt{\textstyle{1\over 2}+y}\ \phi_{(p,s_-,q)}(z),
\cr\cr
p'-3=p-q(\hbox{mod }4).
}\eqno(4.18)
$$
The field of $\N{k2}$ model in the l.h.s. of Eq. (4.17)
has been obtained earlier in Eq. (4.8) in another way.
The field in the l.h.s. of Eq. (4.18) is quite new and
possess conformal dimension
$$
\De^{(k,2)}_{pp'q}
={[(k+4)p-(k+2)q]^2-4\over 8(k+2)(k+4)}+{1\over 2}.
\eqno(4.19)
$$
This field belongs to Ramond sector of the superconformal model.
\par
We can guess that in the case of general $l$ we can obtain
all fields $\phi^{(k,l)}_{pp'q}(z)$ with $1\leq p+q-p'\leq 1+2k$
in a similar way.
\medskip
\pno
{\bf 5. Three-Point Correlation Functions}
\medskip
\pno
It is easy to express structure constants for fields (4.13)
using Eq. (2.2.9). Consider an example. Let $l=2$
and let us try to calculate the constant
$C^{(k,2)}_{221,223,133}$ of the theory $\N{k2}$.
Eq. (4.13) gives four expressions for
$C^{(k,2)}_{221,223,133}$ by structure constants
$C^{(i)}_{(p_1,q_1)(p_2,q_2)(p_3,q_3)}$ of $\M{i}$ models
$$
\eqalignno{
&C^{(k)}_{(2,1)(2,2)(1,2)}C^{(k+1)}_{(1,1)(2,3)(2,3)}=
\textstyle{1\over 2},
&(5.1{\rm a})
\cr
&C^{(k)}_{(2,1)(2,3)(1,2)}C^{(k+1)}_{(1,1)(3,3)(2,3)}=0,
&(5.1{\rm b})
\cr
&C^{(k)}_{(2,2)(2,2)(1,2)}C^{(k+1)}_{(2,1)(2,3)(2,3)}=0,
&(5.1{\rm c})
\cr
&C^{(k)}_{(2,2)(2,3)(1,2)}C^{(k+1)}_{(2,1)(3,3)(2,3)}=
\textstyle{1\over 2}.
&(5.1{\rm d})}
$$
We obtain different values for different representatives! But
we can see that the combinations (5.1b) and (5.1c)
forbidden by fusion rules of $\M{k}$ and
$\M{k+1}$ should be rejected.
Indeed the numerator of the coset construction (1.14)
contains several exemplars of each state of the product
$\N{kl}\cdot$(denominator). It means that each vertex operator
of $\N{kl}$ is represented by several vertices which map
different exemplars. The situation is similar to that in
the bosonic representation of minimal models where each state (or field)
is represented by two states (or fields) and some three-point
correlation functions of bosonic representatives vanish
although respective structure constants are nonzero.
\par
Ultimately, we obtain
$$
C^{(k,2)}_{221,223,133}=\textstyle{1\over 2}.
\eqno(5.2)
$$
\par
In general case we write out
$$
\matrix{
C^{(k,l)}_{p_1p'_1q_1,p_2p'_2q_2,p_3p'_3q_3}
={\displaystyle{
\prod_{i=0}^{l-1}C^{(k+i)}_{(s^{(1)}_i,s^{(1)}_{i+1})
(s^{(2)}_i,s^{(2)}_{i+1})(s^{(3)}_i,s^{(3)}_{i+1})}
}\over\displaystyle{
\prod_{i=1}^{l-1}C^{(i)}_{(t^{(1)}_i,t^{(1)}_{i+1})
(t^{(2)}_i,t^{(2)}_{i+1})(t^{(3)}_i,t^{(3)}_{i+1})} }},
\cr\cr
p'_j=p_j-q_j(\hbox{mod }2l),
}\eqno(5.3)
$$
where $s_i^{(j)}$, $t_i^{(j)}$ are defined by Eqs. (4.11$-$13),
the representation is choosen so that none of structure
constants in (5.3) equal to zero. Structure constants of
minimal models can be adopted from Ref. 12.
\medskip
\pno
{\bf Acknolegments}
\medskip
\pno
I am grateful to V. A. Fateev, Vl. S. Dotsenko, B. L. Feigin,
A. A. Belavin, Ya. P. Pugay, S. E. Parkhomenko and S. V. Kryukov
for useful discussions. I take the opportunity to thank H. D. Toomassian,
V. A. Sadov and S. Piunikhin for many stimulating discussions concerning
different questions of conformal field theory. Besides I want to thank
P.-K. Mitter and Vl. S. Dotsenko for their hospitality in
Laboratoire des Phisique Th\'{e}oretique et Hautes Energies,
Universit\'{e} Pierre et Marie Curie, Paris.
\medskip
\pno
{\bf Appendix A. Quantum groups of minimal models}
\medskip
\pno
Following Sadov\ref{28} we shall show here that quantum group
of the minimal model $\M{k}$ is the usual direct product
$\slq{q(k)}\times\slq{\overline{q(k+1)}}$. We start from
the quantum enveloping algebra obtained by Gomez and Sierra\ref{25}:
$$
\eqalignno{
k_+k_-
&=k_-k_+,
&({\rm A}.1{\rm a})
\cr
k_\pm E_\pm
&=q_\pm^{-1/2}E_\pm k_\pm,\ k_\pm E_\mp=-E_\mp k_\pm,
&({\rm A}.1{\rm b,c})
\cr
k_\pm F_\pm
&=q_\pm^{1/2}F_\pm k_\pm,\ k_\pm F_\mp=-F_\mp k_\pm,
&({\rm A}.1{\rm d,e})
\cr
E_+E_-
&=E_-E_+,\ F_+F_-=F_-F_+,
&({\rm A}.1{\rm f,g})
\cr
E_\pm F_\pm-q_\pm F_\pm E_\pm
&={1-k_\pm^4\over 1-q_\pm^{-1}},
\ E_\pm F_\mp=F_\mp E_\pm,
&({\rm A}.1{\rm h,i})
\cr
q_+
&=q(k),\ q_-=\overline{q(k+1)},
&({\rm A}.1{\rm j})}
$$
with comultiplication
$$
\eqalignno{
\De k_i
&=k_i\otimes k_i,
&({\rm A}.2{\rm a})
\cr
\De E_i
&=E_i\otimes 1+k_i^2\otimes E_i,
&({\rm A}.2{\rm b})
\cr
\De F_i
&=F_i\otimes 1+k_i^2\otimes F_i.
&({\rm A}.2{\rm c})}
$$
It seems that this algebra differs from the direct product by signs
minus in Eqs. (A.1c,e). Let us introduce more usual generators\ref{29}
$X_i^\pm$, $H_i$ by formulae
$$
\eqalignno{
k_\pm
&=q_\pm^{-H_\pm/4}e^{{i\pi\over 2}H_\mp},
&({\rm A}.3{\rm a})
\cr
E_i
&=q_i^{-H_i/4}X_i^+,
&({\rm A}.3{\rm b})
\cr
F_i
&=q_i^{-H_i/4}X_i^-.
&({\rm A}.3{\rm c})}
$$
Substituting (A.3) in (A.1) we obtain immediately
$$
\eqalignno{
\left[X_i^\pm,H_i\right]
&=\mp2X_i^\pm,
&({\rm A}.4{\rm a})
\cr
\left[X_i^+,X_i^-\right]
&={q_i^{H_i/2}-q_i^{-H_i/2}\over q_i^{1/2}-q_i^{-1/2}},
&({\rm A}.4{\rm b})
\cr
[X_+^\aa,X_-^\bb]
&=[X_+^\aa,H_-]=[H_+,X_-^\aa]=0.
&({\rm A}.4{\rm c})}
$$
We see that it is a direct product of two associative
algebras. Substituting (A.3) in (A.2) we see
$$
\eqalignno{
\De H_i
&=H_i\otimes 1+1\otimes H_i,
&({\rm A}.5{\rm a})
\cr
\De X_\pm^\aa
&=X_\pm^\aa\otimes q_\pm^{H_\pm/4}
+q_\pm^{-H_\pm/4}e^{i\pi H_\mp}\otimes X_\pm^\aa.
&({\rm A}.5{\rm b})}
$$
This comultiplication differs from usual one in
$\slq{q(k)}\times\slq{\overline{q(k+1)}}$ by the factor
$\exp(i\pi H_\mp)$ in the second term. To get rid of it we
recall that there are two copies of algebra in Eq. (A.5):
one to the left of sign $\otimes$ and one to the right of it.
In each of them independently we can make any endomorphism of
algebra. Namely, let us make a substitution in the ``left''
algebra
$$
X_\pm^\aa\longmapsto e^{i\pi H_\mp}X_\pm^\aa,
\eqno({\rm A}.6)
$$
and, in the result of comultiplication
$$
\De X_\pm^\aa\longmapsto \left(e^{i\pi H_\mp}\otimes 1\right)
\De X_\pm^\aa.
\eqno({\rm A}.7)
$$
We obtain the usual comultiplication
$$
\De X_\pm^\aa=X_\pm^\aa\otimes q^{H_\pm/4}
+q^{H_\pm/4}\otimes X_\pm^\aa.
\eqno({\rm A}.8)
$$
In terms of conformal blocks the substitution (A.6-7) is no
more than changing signs of some conformal blocks.
\medskip
\pno
{\bf Appendix B. Quantum group $\slq{q(1)}$}
\medskip
\pno
In this Appendix we shall show that quantum group $\slq{q}$
with $q=\exp\left({\2pi\over 3}\right)$ corresponds to
unique intermediate module in any conformal block.
In other words, product of any two irreducible representations
of quantum group algebra is irreducible. Indeed, there are
two irreducible representations $[j]$ with ``spin'' $j=0,{1\over 2}$.
It is evident that
$$
\eqalign{
&[0]\otimes[0]\sim[0],
\cr
&[0]\otimes[\textstyle{1\over 2}]\sim[\textstyle{1\over 2}],
\cr
&[\textstyle{1\over 2}]\otimes[\textstyle{1\over 2}]\sim[0],
}\eqno({\rm B}.1)
$$
which proves the statement. In terms of monodromy
it means that all intermediate states have the same
fractional parts of conformal dimensions and every
monodromy matrix is diagonal.
\medskip
\pno
{\bf Appendix C. Bosonic representation}
\medskip
\pno
In this Appendix we shall sketch some results concerning
bosonization of minimal \ref{10-13} and minimal-like\ref{19,20}
models.
\par
The minimal model $\M{k}$ is described by one free bosonic field
$\varphi(z)$ with correlation function
$$
\langle\varphi(z')\varphi(z)\rangle=-\ln(z'-z),
\eqno({\rm C}.1)
$$
and energy-momentum tensor
$$
T_k(z)=-{1\over 2}:(\partial\varphi)^2:+{i\over\sqrt{2(k+2)(k+3)}}
\partial^2\varphi,
\eqno({\rm C}.2)
$$
which obeys the OPE (1.10) with central charge (1.12). Colons mean
normal ordering, $\partial\equiv\partial/\partial z$. Vertex
operators (3.1) are given by
$$
\eqalign{
\left(\phi^{(k)}_{(p,q)}(z)\right)_{m,n}
&=V_{(p,q)}(z)\prod_{i=1}^m\oint_{C_i}du_iI_+(u_i)
\prod_{j=1}^n\oint_{S_j}dv_jI_-(v_j),
\cr
V_{(p,q)}(z)
&=\ :\exp\left(-i{(k+3)(p-1)-(k+2)(q-1)\over\sqrt{2(k+2)(k+3)}}
\varphi(z)\right):\ ,
\cr
I_+(z)
&=\ :\exp\left(i\sqrt{2{k+3\over k+2}}\ \varphi(z)\right):\
=V_{(-1,1)}(z),
\cr
I_-(z)
&=\ :\exp\left(-i\sqrt{2{k+2\over k+3}}\ \varphi(z)\right):\
=V_{(1,-1)}(z).
}\eqno({\rm C}.3)
$$
Here $I_+(z)$ and $I_-(z)$ are screening fields with conformal
dimension 1. Integration contours\ref{13} are depicted in Fig. 6.
Four-point conformal blocks are given by
$$
\matrix{
\eqalign{
&\Phi_{\{(p,q)\}}\big((p_5,q_5);z\big)
\cr
&=\Big\langle\left(\phi^{(k)}_{(p_1,q_1)}(0)\right)_{0,0}
\left(\phi^{(k)}_{(p_2,q_2)}(z)\right)_{m,n}
\left(\phi^{(k)}_{(p_3,q_3)}(1)\right)_{m',n'}
\left(\tilde{\phi}^{(k)}_{(p_4,q_4)}(\infty)\right)_{0,0}
\Big\rangle,
}\cr\cr
m=\textstyle{1\over 2}(p_1+p_2-p_5-1),
\quad n=\textstyle{1\over 2}(q_1+q_2-q_5-1),
\cr\cr
m'=\textstyle{1\over 2}(p_5+p_3-p_4-1),
\quad n'=\textstyle{1\over 2}(q_5+q_3-q_4-1),
}\eqno({\rm C}.4)
$$
where charge at infinity\ref{10} $\sqrt{2\over(k+2)(k+3)}$
is implicit and the conjugate field is given by
$$
\left(\tilde{\phi}^{(k)}_{(p,q)}(z)\right)_{m,n}
\sim\left(\phi^{(k)}_{(k+2-p,k+3-q)}(z)\right)_{m,n}.
\eqno({\rm C}.5)
$$
Contours in Fig. 6 are not very convenient from the technical
point of view and we shall use another normalization of
conformal blocks\ref{10,11}
$$
\eqalign{
\Phi_{\{(p,q)\}}\big((p_5,q_5);z)
&=\int_0^zdu_1\cdots\int_0^{u_{m-1}}du_m
\int_0^zdv_1\cdots\int_0^{v_{n-1}}dv_n
\cr
&\times\int_1^\infty du'_1\cdots\int_{u'_{m'-1}}^\infty du'_{m'}
\int_1^\infty dv'_1\cdots\int_{v'_{n'-1}}^\infty dv'_{n'}
\cr
&\times\Big\langle V_{(p_1,q_1)}(0)V_{(p_2,q_2)}(z)
V_{(p_3,q_3)}(1)\tilde{V}_{(p_4,q_4)}(\infty)
\cr
&\times\prod_{i=1}^m I_+(u_i)\prod_{j=1}^n I_-(v_j)
\prod_{i=1}^{m'}I_+(u'_i)\prod_{j=1}^{n'}I_-(v'_j)
\Big\rangle.
}\eqno({\rm C}.6)
$$
Coefficients in Eq. (2.1.1) become\ref{11}
$$
\eqalign{
X_{\{(p,q)\}}(p_5,q_5)
&=X_{\{p\}}(p_5;k)\overline{X_{\{q\}}(q_5;k+1)}
\cr
X_{\{p\}}(p_5;k)
&=\prod_{i=1}^ms(i\rho)\prod_{i=1}^{m'}s(i\rho)
\prod_{i=0}^{m-1}{s(a+i\rho)s(b+i\rho)\over s(a+b+i\rho)}
\cr
&\prod_{i=0}^{m'-1}{s(c+i\rho)s(a+b+c+(2m+2n-2-i)\rho)
\over s(a+c+(2m+i)\rho)},
\cr
s(x)
&=\sin\pi x,
\cr
m
&=\textstyle{1\over 2}(p_1+p_2-p_5-1),
\quad m'=\textstyle{1\over 2}(p_5+p_3-p_4-1),
\cr
a
&=-{p_1-1\over k+2},\quad b=-{p_2-1\over k+2},
\quad c=-{p_3-1\over k+2},\quad \rho={1\over k+2}.
}\eqno({\rm C}.7)
$$
The sign of complex conjugation is used to make clearer
the connection with convolution of conformal blocks.
\par
Using (C.6) and (C.7) the following rule for
calculating structure constants can be obtained.\ref{12}
Let us calculate the constants
$$
\eqalign{
A^{(k)(p_3,q_3)}_{(p_1,q_1)(p_2,q_2)}
&=\int_0^1du_1\cdots\int_0^{u_{m-1}}du_m
\int_0^1dv_1\cdots\int_0^{v_{n-1}}dv_n
\cr
&\times\Big\langle V_{(p_1,q_1)}(0)V_{(p_2,q_2)}(1)
\tilde{V}_{(p_3,q_3)}(\infty)
\prod_{i=1}^m I_+(u_i)\prod_{j=1}^n I_-(v_j)
\Big\rangle.
\cr
&m=\textstyle{1\over 2}(p_1+p_2-p_3-1),
\ n=\textstyle{1\over 2}(q_1+q_2-q_3-1).
}\eqno({\rm C}.8)
$$
The constants $A$ take the form\ref{11}
$$
A^{(k)(p_3,q_3)}_{(p_1,q_1)(p_2,q_2)}
=\prod_i\Gamma^{d_i}(a_i),
\eqno({\rm C}.9)
$$
where numbers $d_i$ and $a_i$ can be found in Ref. 11.
Consider quantities
$$
\tilde{A}^{(k)(p_3,q_3)}_{(p_1,q_1)(p_2,q_2)}
=\prod_i\left[{\Gamma(a_i)\over\Gamma(1-a_i)}\right]^{d_i}.
\eqno({\rm C}.10)
$$
Structure constants are given by\ref{12}
$$
\left(C^{(k)}_{(p_1,q_1)(p_2,q_2)(p_3,q_3)}\right)^2
={\tilde{A}^{(k)(p_3,q_3)}_{(p_1,q_1)(p_2,q_2)}
\tilde{A}^{(k)(p_1,q_1)}_{(p_3,q_3)(p_2,q_2)}
\over
\tilde{A}^{(k)(1,1)}_{(p_2,q_2)(p_2,q_2)}}.
\eqno({\rm C}.11)
$$
\par
Now we cite results concerning OPE's of currents which
appear in convolutions of minimal models. Take, for example,
the current $\phi^{(k)}_{(1,s_0)}(z)\phi^{(k+1)}_{(s_0,1)}(z)$,
and consider the operator product expansion
$$
\eqalign{
\phi^{(k)}_{(1,s_0)}(z')\phi^{(k+1)}_{(s_0,1)}(z')
&\cdot\phi^{(k)}_{(p,s)}(z)\phi^{(k+1)}_{(s,q)}(z)
\cr
&=\sum_{s'}(z'-z)^{\aa_{s'}}B^{s'}_{s_0s}(p,q)
\phi^{(k)}_{(p,s')}(z)\phi^{(k+1)}_{(s',q)}(z)+
({\rm decendants}),
}\eqno({\rm C}.12)
$$
where $\aa_{s'}$ are real constants. Note that coefficients
$B^{s'}_{s_0s}(p,q)$ are well defined because all exponents
$\aa_{s'}$ differ by integers. Let us introduce coefficients
$$
\hat{A}^{s'}_{s_0s}(p,q)=KK'\prod_i\Gamma^{d_i}(a_i)
\prod_i\Gamma^{-d'_i}(1-a'_i),
\eqno({\rm C}.13)
$$
where
$$
A^{(k)(p_3,q_3)}_{(p_1,q_1)(p_2,q_2)}=K\prod_i\Gamma^{d_i}(a_i),\quad
A^{(k+1)(p_3,q_3)}_{(p_1,q_1)(p_2,q_2)}=K'\prod_i\Gamma^{d'_i}(a'_i),
\eqno({\rm C}.14)
$$
and $a_i$, $a'_i$ are all arguments of gamma-functions, such that
$$
(k+3)a_i\in{\bf Z},\ (k+3)a'_i\in{\bf Z}.
\eqno({\rm C}.15)
$$
Then
$$
B^{s'}_{s_0s}(p,q)
={\hat{A}^{s'}_{s_0s}(p,q)\hat{A}^{s}_{s_0s'}(p,q)
\over\hat{A}^{1}_{s_0s_0}(1,1)}.
\eqno({\rm C}.16)
$$
Note that
$$
d'_i=d_i,\ a_i+a'_i\in{\bf Z}.
\eqno({\rm C}.17)
$$
\par
Generalizations to more complicated convolutions are evident.
We write out clearly a three-point
correlation function necessary for results of Sec. 3
$$
\eqalign{
\langle\phi_{(p,s,q)}(0,0)
&t^{(n)}_{k+m}(1)\phi_{(p,s,q)}(\infty,\infty)\rangle
\cr
&={(-1)^{n+1}\over 2}{[(k+m+3)s_m-(k+m+2)s_{m+1}]^2-1
\over\sqrt{3(k+m)(k+m+1)(k+m+n+5)(k+m+n+6)}}
\cr
&\times\prod_{i=m+1}^{m+n+1}
{\Gamma\left(s_i-s_{i+1}+1+{s_i+1\over k+i+2}\right)
\Gamma\left(s_{i-1}-s_i+{s_i-1\over k+i+2}\right)
\over\Gamma\left(s_{i-1}-s_i+1+{s_i+1\over k+i+2}\right)
\Gamma\left(s_i-s_{i+1}+{s_i-1\over k+i+2}\right)}.
}\eqno({\rm C}.18)
$$
\par
Bosonic representation of the minimal-like theory $\N{kl}$
is given by three free bosons $\chi(z)$, $\rho(z)$ and
$\varphi(z)$ with nonzero correlation fuctions
$$
\langle\chi(z')\chi(z)\rangle=
\langle\rho(z')\rho(z)\rangle=
\langle\varphi(z')\varphi(z)\rangle=
-\ln(z'-z),
\eqno({\rm C}.19)
$$
and energy-momentum tensor
$$
\eqalign{
T_{k,l}(z)
&=-{1\over 2}:(\partial\chi)^2:-{i\over 2}\partial^2\chi
-{1\over 2}:(\partial\rho)^2:+{1\over 2}\sqrt{l\over l+2}
\ \partial^2\rho
\cr
&-{1\over 2}:(\partial\varphi)^2:+i\sqrt{l\over 2(k+2)(k+l+2)}
\ \partial^2\varphi,
}\eqno({\rm C}.20)
$$
which obey the OPE (1.10) with central charge (1.8).
Screening fields are given by
$$
\eqalign{
I_1(z)
&=\oint_{C_z}{du\over\2pi}:e^{i\chi}:(u)
:\exp\left[-2i\chi+\sqrt{l+2\over l}\ \rho+i\sqrt{2{k+l+2\over l(k+2)}}
\ \varphi\right]:(z),
\cr
I'_1(z)
&=\oint_{C_z}{du\over\2pi}:e^{i\chi}:(u)
:\exp\left[-2i\chi+\sqrt{l\over l+2}\ \rho\right]:(z),
\cr
I_2(z)
&=\oint_{C_z}{du\over\2pi}:\exp\left[i(l+1)\chi-\sqrt{l(l+2)}
\ \rho\right]:(u)
\cr
&\times:\exp\Biggl[-il\chi+(l-1)\sqrt{l+2\over l}\ \rho
-i\sqrt{2{k+2\over l(k+l+2)}}\ \varphi\Biggr]:(z),
}\eqno({\rm C}.21)
$$
contour $C_z$ is a small circle with center $z$. There are four
fields with conformal dimension 0
$$
\eqalign{
Z(z)
&=\oint_{C_z}{du\over\2pi}:e^{i\chi}:(u)
:\exp\left[-i(l+1)\chi+\sqrt{l(l+2)}\ \rho\right]:(z),
\cr
\tilde{Z}(z)
&=\oint_{C_z}{du\over\2pi}
:\exp\left[i(l+1)\chi-\sqrt{l(l+2)}\ \rho\right]:(u):e^{-i\chi}:(z),
\cr
Z'(z)
&=\oint_{C_z}{du\over\2pi}:e^{i\chi}:(u)
:\exp\Bigg[-i(l+1)\chi+(k+2)\sqrt{l+2\over l}\ \rho
\cr
&+i\sqrt{2{(k+2)(k+l+2)\over l}}\ \varphi\Bigg]:(z),
\cr
\tilde{Z}'(z)
&=\oint_{C_z}{du\over\2pi}
:\exp\Bigg[i(l+1)\chi-(k+2)\sqrt{l+2\over l}\ \rho
\cr
&-i\sqrt{2{(k+2)(k+l+2)\over l}}\ \varphi\Bigg]:(u):e^{-i\chi}:(z).
}\eqno({\rm C}.22)
$$
For any field $\phi(z)$ we shall write
$$
W\phi(z)=\oint_{C_z}{du\over\2pi}{Z(u)\over u}\phi(z),
\eqno({\rm C}.23)
$$
and similar for $\tilde{Z}$, $Z'$, $\tilde{Z}'$.
\par
Vertex operators are given by\ref{20}
$$
\eqalign{
\left(\phi^{(k,l)}_{pp'q}(z)\right)_{mm'n}
&=V_{pp'q}(z)\prod_{i=1}^m\oint_{C_i}du_iI_1(u_i)
\prod_{i=1}^{m'}\oint_{C'_i}du'_iI'_1(u'_i)
\prod_{i=1}^n\oint_{S_i}dv_iI_2(v_i);
\cr
V_{pp'q}(z)
&=W^N\psi^{N,0}_{pp'q}(z)\hbox{ or }
\tilde{W}^{-N}\psi^{N,0}_{pp'q}(z),
\cr
\hbox{ if }-p'+1\leq
&q-p-2lN\leq-p'+1+2l,\ -p+1\leq q-p'\leq-p+1+2k;
\cr
V_{pp'q}(z)
&=W^{\prime N'}\psi^{0,N'}_{pp'q}(z)\hbox{ or }
\tilde{W}^{\prime-N'}\psi^{0,N'}_{pp'q}(z),
\cr
\hbox{ if }-p+1\leq
&q-p'-2kN'\leq-p+1+2k,\ -p'+1\leq q-p\leq-p'+1+2l;
\cr
\psi^{N,N'}_{pp'q}(z)
&=\psi_{p-2(k+2)N,p'-2(l+2)N',q-2(k+l+2)(N+N')}(z),
\cr
\psi_{pp'q}(z)
&=f_{p+p'-q-1}(z):\exp\left[{(l+2)(q-p)-l(p'-1)
\over 2\sqrt{l(l+2)}}\rho\right.
\cr
&\left.-i{(k+l+2)(p-1)-(k+2)(q-1)\over\sqrt{2l(k+2)(k+l+2)}}\varphi
\right]:(z),
\cr
f_{2n}(z)
&=\cases{
:e^{in\chi}:(z)& if $n\geq 0$,
\cr
\oint_{C_z}{du\over\2pi}:e^{i\chi}:(u):e^{i(n-1)\chi}:(z)
& if $n\leq 0$.}
}\eqno({\rm C}.24)
$$
Conformal dimensions are given by
$$
\matrix{
\eqalign{
\De_{pp'q}
&={[(k+l+2)p-(k+2)q]^2-l^2\over 4l(k+2)(k+l+2)}
+{p^{\prime 2}-1\over 4(l+2)}-{t^2\over 4l}
\cr
&+\textstyle{1\over 2}(t-p'+1)\theta(t-p'+1),
}\cr\cr
q-p-t\in 2l{\bf Z},\ -p'+1\leq t\leq-p'+1+2l,
\ -p+1\leq q-p'\leq-p+1+2k.
}\eqno({\rm C}.25)
$$
If $-p+1>q-p'$ or $q-p'>-p+1+2k$, we can use the same formula
with substitution $k\longleftrightarrow l$,
$p\longleftrightarrow p'$.
\par
Note that these vertices do not exaust all fields primary with respect
to Virasoro algebra. Nevertheless, every field of the theory
is contained in the Fock space of one of these bosonic vertices.
\par
At last we write out some bosonic three-point correlation
functions which are used together with (C.16)-like equations
in the proof of Eqs. (3.14), (3.17), (3.20) and (3.22):
$$
\eqalign{
\Big\langle\phi_{131}(0)\phi_{131}(1)
&\tilde{\phi}_{131}(\infty)
\int_0^1duI'_1(u)\Big\rangle
=2{2k+l+4\over l+4}{\Gamma^2\left(-{2\over l+2}\right)
\over\Gamma\left(-{4\over l+2}\right)},
\cr
\Big\langle\phi_{131}(0)\phi_{131}(1)
&\tilde{\phi}_{111}(\infty)
\int_0^1du_1\int_0^{u_1}du_2I'_1(u_1)I'_1(u_2)\Big\rangle
\cr
&=-2{k(k+l+4)\over l+4}{\Gamma\left({2\over l+2}\right)
\Gamma\left(-{2\over l+2}\right)\Gamma^2\left(-{1\over l+2}\right)
\over\Gamma\left({1\over l+2}\right)
\Gamma\left(-{3\over l+2}\right)},
\cr
\Big\langle\phi_{113}(0)\phi_{113}(1)
&\tilde{\phi}_{113}(\infty)
\int_0^1duI_2(u)\Big\rangle
=-2{k-l\over k+l}{\Gamma^2\left({2\over k+l+2}\right)
\over\Gamma\left({4\over k+l+2}\right)},
\cr
\Big\langle\phi_{113}(0)\phi_{113}(1)
&\tilde{\phi}_{111}(\infty)
\int_0^1du_1\int_0^{u_1}du_2I_2(u_1)I_2(u_2)\Big\rangle
\cr
&=-2{kl\over k+l}{\Gamma\left(-{2\over k+l+2}\right)
\Gamma\left({2\over k+l+2}\right)\Gamma^2\left({1\over k+l+2}\right)
\over\Gamma\left(-{1\over k+l+2}\right)
\Gamma\left({3\over k+l+2}\right)}.
}\eqno({\rm {\rm C}}.26)
$$
\vfill\eject
\par\noindent
{\bf References}
\par\noindent
\pno
1. P.Goddard, A.Kent and D.Olive, $Phys$. $Lett$. {\bf 152}, 88 (1985)
\pno
2. P.Goddard, A.Kent and D.Olive, $Commun$. $Math$. $Phys$.
{\bf 103}, 105 (1986)
\pno
3. A. A. Belavin, A. M. Polyakov and A. B. Zamolodchikov, $Nucl$. $Phys$.
{\bf B241},
\par
333 (1984)
\pno
4. K.Gawedzki, preprint IHES/P/89/53 (August 1989)
\pno
5. M. R. Douglass, preprint CALT-68-1453 (1987)
\pno
6. M. B. Halpern and N. Obers, preprint LBI-32619, USB-PTH-92-24,
BONN-
\par
HE-92/21, hep-th/9207071 (July 1992)
\pno
7. M. Yu. Lashkevich, preprint LANDAU-92-TMP-1,
hep-th/9301094 (October
\par
1992)
\pno
8. E. Witten, $Commun$. $Math$. $Phys$. {\bf 92}, 455 (1984)
\pno
9. V. G. Knizhnik and A. B. Zamolodchikov, $Nucl$. $Phys$. {\bf B247}, 83
(1984)
\pno
10. Vl. S. Dotsenko and V. A. Fateev, $Nucl$. $Phys$.
{\bf B240 [FS12]}, 312 (1984)
\pno
11. Vl. S. Dotsenko and V. A. Fateev, $Nucl$. $Phys$.
{\bf B251 [FS13]}, 691 (1985)
\pno
12. Vl. S. Dotsenko and V. A. Fateev, $Phys$. $Lett$. {\bf B154}, 291 (1985)
\pno
13. G. Felder, $Nucl$. $Phys$. {\bf B317}, 215 (1989)
\pno
14. H. Eichenherr, $Phys$. $Lett$. {\bf B151}, 26 (1985)
\pno
15. M. A. Bershadsky, V. G. Knizhnik and M. G. Teitelman, $Phys$. $Lett$.
{\bf B151},
\par
31 (1985)
\pno
16. D. Freedan, Z. Qiu and S. Shenker, $Phys$. $Lett$. {\bf B151}, 37 (1985)
\pno
17. J. Bagger, D. Nemeschansky and S. Yankielowicz,
$Phys$. $Rev$. $Lett$. {\bf 60}, 389
\par
(1988)
\pno
18. D. Kastor, E. Martinec and Z. Qiu, $Phys$. $Lett$. {\bf B200}, 434 (1988)
\pno
19. P. Bouwknegt, J. McCarthy and K. Pilch, $Prog$. $Theor$. $Phys$. $Supp.$
No.102,
\par
67 (1990)
\pno
20. M. Yu. Lashkevich, $Int$. $J$. $Mod$. $Phys$. {\bf A7}, 6623 (1992)
\pno
21. M. Yu. Lashkevich, to be published in $Mod$. $Phys$. $Lett$. $A$
\pno
22. M. Yu. Lashkevich, preprint LANDAU-92-TMP-1,
hep-th/9301093 (Decebmer
\par
1992); to be published in $Mod$. $Phys$. $Lett$. $A$
\pno
23. A. B. Zamolodchikov and V. A. Fateev,
$Sov$. $Phys$. $JETP$ {\bf 62}, 215 (1985)
\pno
24. G. Felder, J. Fr\"olich and G. Keller, $Commun$. $Math$. $Phys$.
{\bf 130}, 1 (1990)
\pno
25. C. Gomez and G. Sierra, $Nucl$. $Phys$. {\bf B352}, 791 (1991)
\pno
26. C. Ramirez, H. Ruegg and M. Ruiz-Altaba, $Nucl$. $Phys$.
{\bf B364}, 195 (1991)
\pno
27. J.-L. Gervais, $Commun$. $Math$. $Phys$. {\bf 130}, 257 (1990)
\pno
28. V. A. Sadov, private communication
\pno
29. A. N. Kirillov and N. Yu. Reshetikhin, LOMI preprint E-9-88 (April 1988)
\end